\documentclass[preprint,journal]{vgtc}       

\usepackage{mathptmx}
\usepackage{graphicx}
\usepackage{times}
\usepackage{xspace}
\usepackage{xcolor}
\usepackage{enumitem}
\usepackage{booktabs} 
\usepackage{listings}
\usepackage{makecell}
\usepackage[normalem]{ulem}
\usepackage{kotex}
\usepackage{multirow}
\usepackage{wrapfig}
\usepackage{lipsum}
\usepackage{amsfonts}

\usepackage[bookmarks,backref=true,linkcolor=black]{hyperref} 
\hypersetup{
  pdfauthor = {},
  pdftitle = {},
  pdfsubject = {},
  pdfkeywords = {},
  colorlinks=true,
  linkcolor= black,
  citecolor= black,
  pageanchor=true,
  urlcolor = black,
  plainpages = false,
  linktocpage
}

\newcommand{\needEffectiveHyper}{\hyperlink{needEffectiveHyper}{N1\xspace}}
\newcommand{\needUnderstandMethod}{\hyperlink{needUnderstandMethod}{N2\xspace}}
\newcommand{\needSelectModel}{\hyperlink{needSelectModel}{N3\xspace}}
\newcommand{\goalEffectiveHyper}{\hyperlink{goalEffectiveHyper}{G1\xspace}}
\newcommand{\goalUnderstandMethod}{\hyperlink{goalUnderstandMethod}{G2\xspace}}
\newcommand{\goalSelectModel}{\hyperlink{goalSelectModel}{G3\xspace}}
\newcommand{\goalIterativeOpt}{\hyperlink{goalIterativeOpt}{G4\xspace}}

\newcommand{\toolname}{HyperTendril\xspace}
\newcommand{\methodLevel}{search method-level\xspace}
\newcommand{\hyperparamLevel}{hyperparameter-level\xspace}
\newcommand{\modelLevel}{model-level\xspace}

\newcommand{\controlPanel}{Control panel\xspace}
\newcommand{\mainModule}{AutoML analytics\xspace}
\newcommand{\optimizationOverview}{Optimization overview\xspace}
\newcommand{\searchSpaceView}{Search space overview\xspace}
\newcommand{\modelAnalysisView}{Model analysis view\xspace}
\newcommand{\explorationOverview}{Exploration overview\xspace}
\newcommand{\experimentTable}{Selected experiments\xspace}
\newcommand{\hyperImpView}{Hyperparameter importance view\xspace}

\newcommand{\footnoteTendril}{\footnote{The name of the system comes from a vine plant stem called Tendril, which implies the design goals of our system.}\xspace}
\newcommand{\footnoteNaver}{\footnote{
NAVER Co., Ltd. is the largest web search engine in South Korea and a global ICT brand that provides services such as LINE messenger and webtoon.}
}
\newcommand{\footnoteDataDrivenDocument}{\footnote{https://d3js.org/}\xspace}
\newcommand{\footnoteReact}{\footnote{https://reactjs.org/}\xspace}
\newcommand{\footnoteFanova}{\footnote{https://github.com/automl/fanova}\xspace}

\onlineid{1105}

\vgtccategory{Application/design study}

\vgtcinsertpkg

\title{\toolname: Visual Analytics for User-Driven Hyperparameter Optimization of Deep Neural Networks}

\author{Heungseok Park, Yoonsoo Nam, Ji-Hoon Kim$^\ast$, and Jaegul Choo$^\ast$}
\authorfooter{
\item
 Heungseok Park, Yoonsoo Nam, and Ji-Hoon Kim are with Clova AI Research, NAVER Corporation. E-mail: {heungseok.park, yoonsoo.nam, genesis.kim}@navercorp.com.
\item
 Jaegul Choo is with KAIST. E-mail: jchoo@kaist.ac.kr.

 *Corresponding author.
}

\shortauthortitle{Park \MakeLowercase{\textit{et al.}}: \toolname}

\abstract{
To mitigate the pain of manually tuning hyperparameters of deep neural networks, automated machine learning (AutoML) methods have been developed to search for an optimal set of hyperparameters in large combinatorial search spaces.
However, the search results of AutoML methods significantly depend on initial configurations, making it a non-trivial task to find a proper configuration. 
Therefore, human intervention via a visual analytic approach bears huge potential in this task.
In response, we propose \toolname, a web-based visual analytics system that supports user-driven hyperparameter tuning processes in a \textit{model-agnostic} environment.
\toolname takes a novel approach to effectively steering hyperparameter optimization through an \textit{iterative, interactive} tuning procedure that allows users to refine the search spaces and the configuration of the AutoML method based on their own insights from given results. 
Using \toolname, users can obtain insights into the complex behaviors of various hyperparameter search algorithms and diagnose their configurations.
In addition, \toolname supports variable importance analysis to help the users refine their search spaces based on the analysis of relative importance of different hyperparameters and their interaction effects. We present the evaluation demonstrating how \toolname helps users steer their tuning processes via a longitudinal user study based on the analysis of interaction logs and in-depth interviews while we deploy our system in a professional industrial environment.

}


  \teaser{
 \centering
  \includegraphics[width=15.5cm]{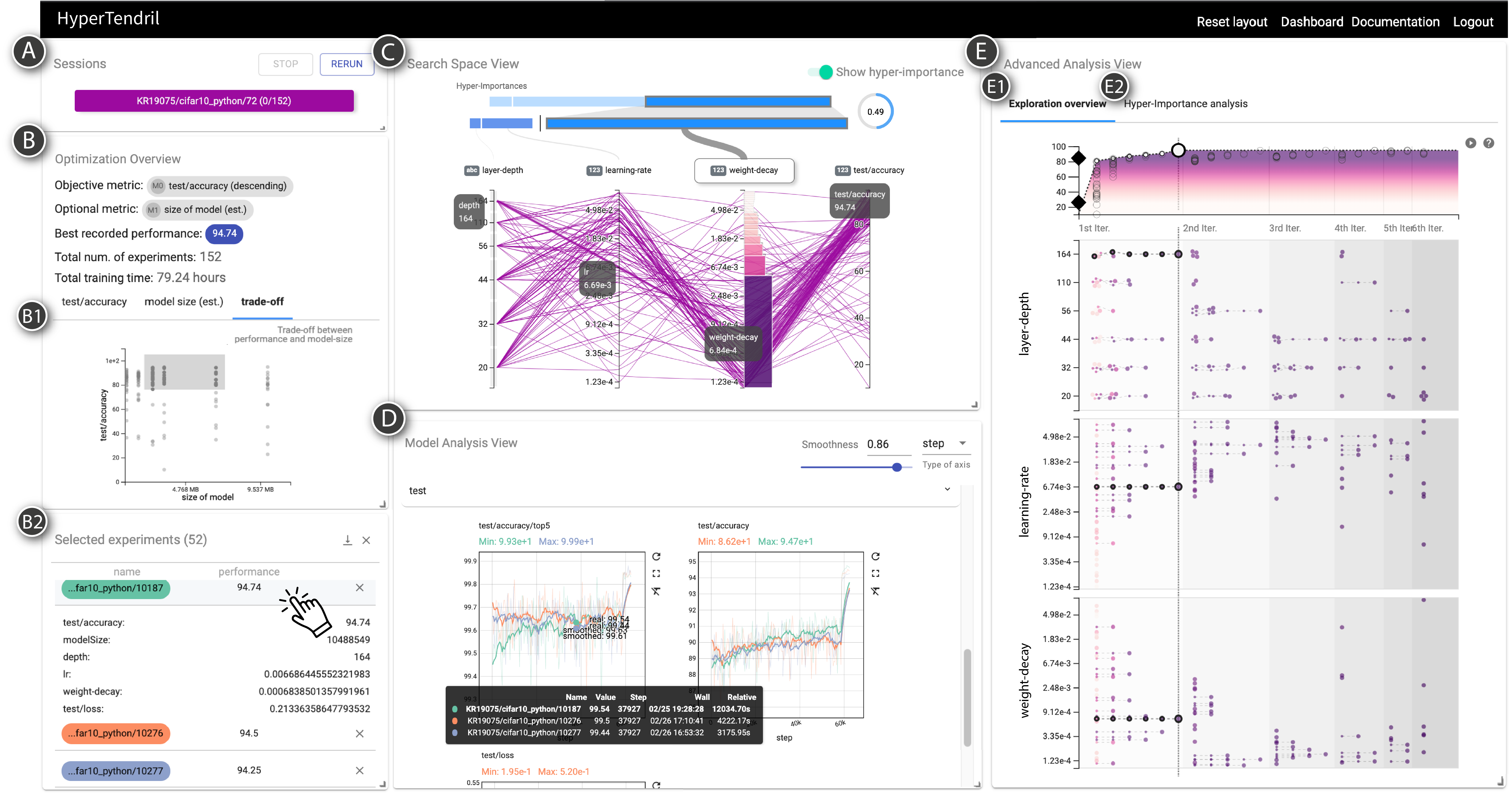}
  \vspace*{-2mm}
  \caption{
  Overview of \toolname that supports user-driven AutoML processes. This example involves the three hyperparameters, e.g., the number of layers, learning rate, and weight decay in the ResNet architecture, using a Bayesian Optimization and HyperBand (BOHB) search method. (C) \searchSpaceView, (D) \modelAnalysisView, and (E1) \explorationOverview components shows the model details selected in (B2) \experimentTable panel. The weight decay hyperparameter is activated in (C) \searchSpaceView, and its effective range is highlighted in the parallel coordinates.
  }
  \label{fig:automlVisTeaser}
  \vspace{-5pt}
  }

\begin{document}

\maketitle
\section{Introduction} 
As deep neural networks evolve with highly modular architectures and advanced optimization methods, an increasing number of hyperparameters are involved. Typically, these hyperparameters need to be optimized either entirely by hand or in a semi-automated manner, requiring significant human efforts and computational resources~\cite{bergstra2012random, goodfellow2016deep}. This issue often hinders researchers and practitioners from finding their optimal settings, left with sub-optimal model performances. Therefore, the methods and the user interfaces for automated hyperparameter optimization (HyperOpt) have emerged as a critical task.

To handle such a task, various optimization methods have been proposed in the machine learning community, say, using a sequential model~\cite{bergstra2011algorithms, shahriari2015taking}, a genetic algorithm~\cite{jaderberg2017population, young2015optimizing}, and a bandit algorithm~\cite{falkner2018bohb, li2017hyperband}.
These studies contributed to developing efficient search methods by sampling hyperparameters based on prior observations, allocating computing resources to potentially promising models (i.e., excluding poor models at an early stage), or combining these approaches.
However, these methods still require considerable time and effort to run, so automated machine learning (AutoML) systems~\cite{golovin2017google, liaw2018tune, moritz2017ray, tsirigotis2018orion} have been developed to help practitioners conveniently optimize their models by providing interfaces for various HyperOpt methods. These systems have numerous advantages including parallelism, early stopping, and ease of use, which can significantly improve efficiency in terms of resource utilization and user experience.

Although these approaches help practitioners effectively optimize their models, they often require delicate configuration settings until obtaining satisfactory results.
For example, the evolutionary optimization algorithms need to carefully set the population size in advance, which determines the number of individuals to generate in each generation, before starting an optimization process, since the convergence behaviors and final results can vary significantly.
In addition, proper settings can differ depending on the task being applied, so they still need to spend considerable time to maximize the potential of the application method. 
Due to the absence of the approaches to arranging a HyperOpt configuration, it is a common practice to go through numerous manual trials with different configurations, with no clear guidance.

In order to alleviate the pain of tedious processes, human intuition for AutoML results, such as the behavior of search algorithms, the effect of optimization algorithm setting, and the characteristics of hyperparameters, should be accompanied.
Thus, effective and efficient human intervention is critical during the HyperOpt process, which necessitates a visual analytics system that can leverage human insights to steer the optimization process in a user-driven manner.
To this end, we propose \toolname (Fig.~\ref{fig:automlVisTeaser}), a web-based visual analytics system that supports HyperOpt tasks, where users can effectively perform HyperOpt through an \textit{iterative} and \textit{interactive} tuning procedure, allowing them to fine-tune the optimal hyperparameters based on their domain knowledge and insights obtained from the previous results.
In detail, \toolname helps the users progressively refine their search spaces by explicitly highlighting relevant hyperparameters and the promising ranges to explore further, based on a quantitative analysis on the objective model performance (e.g., a test accuracy) (Fig.~\ref{fig:automlVisTeaser}(C)).
In addition, \toolname visualizes the exploration history of search algorithms for each search space (Fig.~\ref{fig:automlVisTeaser}(E1)) so that users can visually understand the complex behavior of the used algorithm and compare the differences between the algorithm configurations, enabling them to diagnose and adjust it to their own tasks.

To demonstrate the utility of the proposed system, we deploy our system in an industrial-scale environment and conduct an evaluation focusing on how the visual analytics assists users in steering their tuning processes via a longitudinal user study with interaction log analysis and in-depth interviews with professional users.

The main contributions of \toolname are as follows: 
\begin{itemize}
    \setlength\itemsep{-0.08em}
    \item A novel visual representation that visualizes the exploration history of HyperOpt algorithms, which facilitates understanding of complex behavior of the search algorithms and diagnosis of the algorithm configurations with \textit{algorithm-agnostic} support.
    \item A novel approach to effectively steer users' HyperOpt processes by guiding on the refining search spaces based on quantitative analysis of hyperparameter importance.
    \item We demonstrate a user study to show how our visualization and approach work in an AI research company at an industrial scale.
\end{itemize}

\section {Related Work}
This section discusses recent hyperparameter optimization systems with their visualization modules and visual analytics studies related to refining deep neural networks.

\subsection{Visual Analytics for Hyperparameter Optimization}
Various data exploration systems have been developed to visualize high-dimensional search spaces with parallel coordinates~\cite{inselberg1987parallel} for the analysis of HyperOpt result, by showing the relationships between hyperparameters and performance of a model. 
Golovin et al.~\cite{golovin2017google} developed Google Vizier, an interactive visualization for HyperOpt used internally in Google. They designed dashboard-style interfaces, enabling users to manage and monitor optimization process. Vizier supports a parallel coordinates plot for analyzing the hyperparameters influencing model performance.
Other studies and projects~\cite{akiba2019optuna, web:comet, liaw2018tune, liu2019auptimizer, web:nni, tsirigotis2018orion} for HyperOpt similarly utilize the parallel coordinates, which can be considered as a standard visualization technique for the HyperOpt task.
Even if these existing studies contribute to the analysis and monitoring of optimization results by integrating with their AutoML systems, they lack a consideration of the visual analytics system that leverages human insights required to effectively steer the HyperOpt process in a user-driven manner.

Meanwhile, visualization systems that consider the human-in-the-loop environment have been devised in relation to HyperOpt tasks.
Li et al.~\cite{lihypertuner} studied an empirical hyperparameter search process with practitioners in a software company and described a practical workflow for the task. They developed HyperTuner, allowing users to initialize their HyperOpt processes and analyze a batch of experiments in small multiples of scatter plots showing the effects of each hyperparameter on model performance.
AutoAIViz~\cite{weidele2020autoaiviz} extends the visualization for HyperOpt task to an entire model building pipeline visualization, by utilizing a conditional parallel coordinates design. AutoAIViz is tightly connected with the backend platform~\cite{wang2020autoai} to allow users to directly steer the pipeline optimization.
Wang et al.~\cite{wang2019atmseer} derived a workflow with key decisions during the use of AutoML and developed an open-source visual analytics system called ATMSeer.
ATMSeer provides a multi-granularity visualization of model selection and hyperparameter tuning, enabling users to monitor the process and intervene in the middle of the process to adjust their search spaces in real-time, by tightly integrating with its backend framework called ATM~\cite{swearingen2017atm}.

%
These visualization systems work as powerful data exploration tools for the HyperOpt task and also increase the transparency of the process by leveraging human insights. However, they did not consider two important aspects. 
First, they did not explicitly guide which hyperparameters are important nor which hyperparameter ranges are promising for further exploration.
Consequently, the refinement of search spaces solely depends on the users' intuition. The explicit guidance on effective hyperparameters can assist users in discovering the sweet spot of the search spaces and increase search efficiency.
Second, they did not consider the importance of configuring and diagnosing various HyperOpt algorithms that can have a significant impact on search results. 
Specifically, ATMSeer demonstrated that the system can reveal the bias of the used search algorithm, by comparing the histograms of the search results.
However, from the perspective of non-expert users for AutoML, it is difficult to grasp the reasons for this phenomenon, since the visualization was not designed to present the inner workings of the AutoML algorithm.
Consequently, the transparency and controllability of the AutoML method still remain low since it is difficult to figure out how to configure it to fit their tasks when AutoML produces unreliable results.

Our approach assists users in refining the search space via a quantitative analysis method for measuring hyperparameter importance and visual representation of the analysis results, which provide users with (1) the relative importance of different hyperparameters with the guidance of promising ranges which is worth exploring further. In addition, we provide a visual representation that visualizes the exploration history of hyperparameter search algorithms, which facilitates (2) the understanding of the nature of search algorithms and the diagnosis of the given configuration.

\begin{figure*}[ht!]
\centering
\includegraphics[width=0.9\textwidth]{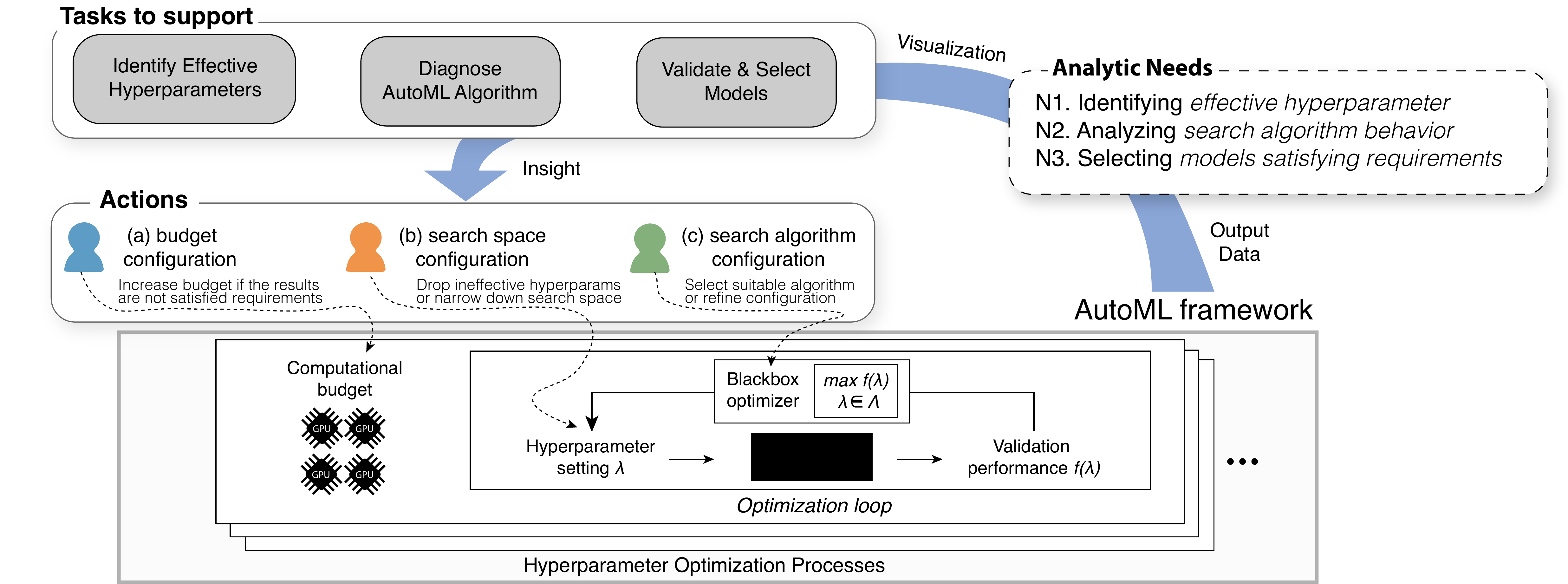}
\vspace*{-3mm}
\caption{Illustration of practical analytic needs and hyperparameter optimization workflow, identified from interviews with ten practitioners.
There are three common analytic needs from the practitioners, and the top left shows tasks in which a visual analytics approach can assist them.
}
\label{fig:automl_workflow}
\vspace{-18pt}
\end{figure*}


\subsection{Visual Analytics for Refining Deep Neural Networks}
Similar to the internal parameters of deep learning models, hyperparameters can have a significant impact on model performance and robustness, but their optimization is difficult to achieve using differential techniques. Therefore, a separate fine-tuning process is still required. In response, recent research has focused on the interactive workflow for the process provided by visual analytics systems~\cite{endert2017state}. These needs for targeted analysis and design of visual analytics for deep learning have been summarized as an exploratory workflow~\cite{sacha2018vis4ml}.

Numerous visual analytics systems are mainly focused on supporting workflow phases that correspond to the model training preparation and model evaluation, allowing practitioners to understand the effects of hyperparameters and iteratively improve it.
TensorFlow Playground~\cite{smilkov2017direct}, ShapeShop~\cite{hohman2017shapeshop}, and ReVACNN~\cite{chung2016re} provide a straightforward analysis system for neural network improvement by allowing the user to directly select the hyperparameters corresponding to the model training preparation stage. ML-o-scope~\cite{Bruckner:EECS-2014-99}, explAIner~\cite{spinner2019explainer}, and DeepEyes~\cite{pezzotti2018deepeyes} introduced a time-lapse engine to investigate the model’s learning dynamics, a pipeline framework for comprehensive model analysis, and incremental model development through activation heatmap, respectively. These systems suggested that the provided analysis tool can help to refine the deep learning model through reinforcement of the model evaluation stage.

From an ontological perspective~\cite{sacha2018vis4ml}, the conventional visual analysis systems are similar to our system that supports hyperparameter analysis and comparative analysis of deep learning models.
However, for quality and result analysis phase, while existing studies only support model performance comparison analysis, our proposed system additionally provides configuration diagnosis and behavior analysis of hyperparameter search algorithms so that users can identify effective hyperparameters through a more optimized search algorithm. 
Therefore, even if the workflow has similar ontology pathway components, the user behavior and model search performance could be more dynamic and efficient than the previous work.

\section {Identification of Analytic Needs in Hyperparameter Optimization Task}~\label{sec:analytics_needs}
To learn how practitioners typically work on the hyperparameter optimization process and deal with resulting problematic issues, and how a visualization domain can assist the tasks and reduce the issues, we first conducted semi-structured interviews with ten machine learning practitioners in the NAVER company\footnoteNaver.
Based on user interviews, we established a common hyperparameter optimization workflow when optimizing their models.
Following the analysis, we present three key analytic needs that a visual analytics approach can support.
These findings guide our discovery of design goals that we aim to address.

\subsection{Background: Optimizing Deep Learning Models with AutoML frameworks}
Hyperparameter optimization (HyperOpt) refers to finding a set of optimal hyperparameter values that have to be set before performing the training of machine learning models. The optimal value of the hyperparameters here is the hyperparameter value that maximizes the performance (i.e., minimizes error rate) of the trained model.
For example, learning rates, training batch sizes, batch normalization, and weight decay are considered as typical hyperparameters in training a deep learning model.
In addition, factors that determine the structure of the deep learning model can also be considered as hyperparameters, such as the number of layers and the size of the convolution filters.

Meanwhile, HyperOpt is performed through an optimization loop, as shown at the bottom of Fig.~\ref{fig:automl_workflow}.
If we let $\lambda$ as a hyperparameter setting of a deep neural network $A$, the optimization loop iteratively samples different $\lambda$ ($\lambda$ $\in$ $\Lambda$) that maximizes the validation performance $f(\lambda)$.
Contrary to the approach in optimizing neural networks, the gradient cannot be computed in this loop, and the loop has no access to any other information about $f$. 
Therefore, $\lambda$ is sampled randomly or through an educated guess from the prior results of given search spaces $\Lambda$.
In order to sample $\lambda$ efficiently, various optimization methods have been proposed in the machine learning community with different approaches, such as sequential model-based~\cite{bergstra2011algorithms, shahriari2015taking}, genetic algorithm-based~\cite{jaderberg2017population}, and bandit-based~\cite{falkner2018bohb, li2017hyperband} optimization.

Most AutoML frameworks~\cite{akiba2019optuna, web:comet, golovin2017google, kim2018chopt, liaw2018tune, liu2019auptimizer, web:nni, swearingen2017atm} are usually designed to generate deep learning models of multiple hyperparameter combinations based on given search spaces and return models with high performance. The systems allow users to configure various settings to incorporate human knowledge and improve search efficiency.
In order to find the optimal hyperparameters through AutoML framework, three key configurations should be established by users at an initial stage in general: (1) hyperparameter candidates to search and their search ranges to explore, (2) a computational budget to use, and (3) a HyperOpt algorithm to perform search and its configuration.
\subsection{Workflow of Hyperparameter Optimization Process}~\label{sec:optimization_workflow}
In order to identify general workflows of the HyperOpt process and analytic needs of real users, we conducted interviews with those who have used the AutoML system launched as an internal beta service since January 2019.
The participants were asked to describe their practices when using the AutoML system, such as how they configure the HyperOpt settings and interpret the results.
The interviews have shown us how they optimize the hyperparameters of their models in general, leading us to identifying the three analytic needs (N1-N3) of visual analytics required during the HyperOpt process.
Following the analysis, we summarize and illustrate the overall workflow of the HyperOpt process with the identified needs that can be supported by visual analytics, as shown in Fig.~\ref{fig:automl_workflow}.

\begin{enumerate}[label={\textbf{N\arabic*.}}, wide, labelwidth=!, labelindent=0pt]
    \item
    \textbf{\hypertarget{needEffectiveHyper}{Identify effective hyperparameters.}}
    As described before, AutoML frameworks usually allow users to configure hyperparameter search spaces as a preset before performing the process.
    Interviewees stated that they usually check the hyperparameter values of the best $k$ ($k<10$, usually) models from the prior results and then narrow down the search spaces by exploring the neighborhood of best hyperparameters with a larger computational budget.
    However, they are not confident about whether the refined search spaces would yield better models. In addition, they want to understand the impact of each hyperparameter on the model performance so that they can increase search efficiency by dropping ineffective hyperparameters from the candidates or narrowing them down to the subspace where a better model can be found.

    \item
    \textbf{\hypertarget{needUnderstandMethod}{Understand and diagnose search algorithms.}}
    Those interviewees, who have used various search algorithms to tune the hyperparameters of their models, are concerned about the configuration of HyperOpt algorithms (e.g., how much population size or survivor rate of the population should be set when using an evolutionary-based optimization algorithm~\cite{bergstra2011algorithms, jaderberg2017population}), since the results can vary depending on the values of the algorithm configurations. 
    In addition, numerous machine learning engineers and experts, even though they had experience with deep learning models, were unfamiliar with various HyperOpt algorithms.
    They said that it was difficult to understand how each search algorithm works and to obtain the configuration value suited for their tasks.
    They cared about the detailed values of the configuration and expressed the necessity of visual analytics for the diagnosis of various algorithms and their configurations.
    
    \item
    \textbf{\hypertarget{needSelectModel}{Select models with user requirements.}}
    AutoML returns the model with the best performance score (e.g., a test accuracy) but does not consider other aspects of the model. However, in practice, model features other than the main performance have to be considered for model selection. For example, some users developing edge-device-related services wanted to ensure that the best-performing model is light-weight to deploy on small devices and to understand which hyperparameters affect the model size. The other users wanted to validate the training process of the model by checking whether the loss function value is adequately low and saturated.
    In addition, if the output model did not satisfy the requirements, they carried out the subsequent HyperOpt processes until a good trade-off point between requirements and performance were found.

    %
    %
\end{enumerate}
A major finding from the conducted interviews is that the HyperOpt process does not end in a single trial.
Due to the main needs (N1-N3) and other auxiliary reasons such as the lack of computing resources, users iteratively performed the HyperOpt process multiple times until obtaining a satisfactory model.
Therefore, even if users take advantage of a HyperOpt framework, they are always in the middle of the optimization loop and have to make a decision on subsequent actions for better results based on the insights gained from the prior results, as illustrated in Fig.~\ref{fig:automl_workflow}.

In summary, the analytic needs and workflow we acquired through the interviews were in line with those presented by Li et al.~\cite{lihypertuner} and Wang et al.~\cite{wang2019atmseer}. However, previous studies did not consider the users' lack of experience with HyperOpt algorithms, which limited users' understanding and means for applying prior knowledge.
It motivated us to develop a more advanced design that does not only improve model performance but also aid the diagnosis of the algorithm’s process.


\section{Design Goals}~\label{sec:design goals}
In this section, we highlight and formalize the primary analytic needs discussed earlier in Section~\ref{sec:analytics_needs} with key design goals that \toolname aims to support.
We label the four goals as G1 - G4.

\begin{enumerate}[label={\textbf{G\arabic*.}}, wide, labelwidth=!, labelindent=0pt]
    \item
    \textbf{\hypertarget{goalEffectiveHyper}{Qualitative and quantitative analysis of effective hyperparameters.}} 
    We aim to support the explicit guidance on effective hyperparameters in both a qualitative and a quantitative manner, by measuring the importance of hyperparameters and visualizing the results effectively (\needEffectiveHyper).
    In addition, we aim to provide an overview interface of the configured search spaces for refining them based on the quantitative analysis results.
    \item
    \textbf{\hypertarget{goalUnderstandMethod}{Effective visual representation for understanding and diagnosing various search algorithms.}}
    The behavior and configuration of various search algorithms are complex and diverse, and the hyperparameter search space of deep learning models is usually high-dimensional.
    Therefore, designing a visual representation for understanding the exploration process of the algorithms can be challenging, and none of the previous studies attempted it.
    We aim to visualize the exploration history of search algorithms in an \textit{algorithm-agnostic} manner by using multiple coordinated search history view of each hyperparameter. It will let the users understand the algorithm behavior on their tasks and choose the proper configurations for the algorithms (\needUnderstandMethod). 
    In addition, we aim to support a monitoring view for the performance of search algorithms so that the users can quickly diagnose whether the algorithm is consistently improving their model performances over time. 

    \item
    \textbf{\hypertarget{goalSelectModel}{Interface for filtering and analyzing models with various perspectives.}}
    In order to support model selection and analysis from the vast number of trained models, it is desirable to filter particular models with user requirements along with detailed information for them (\needSelectModel).
    In general, automated HyperOpt produces a number of deep learning models with different hyperparameter combinations. Visualizing every detail of optimization results can overwhelm users when selecting their desired models.
    Therefore, we aim to present the overview of the optimization results, allowing users to filter the particular models by their desired attributes and drill down to the detailed analysis on demand, by tightly integrating with the overview component.

    \item
    \textbf{\hypertarget{goalIterativeOpt}{Interactive and effective interface for iterative optimization process.}}
    As described in Section~\ref{sec:optimization_workflow}, the hyperparameter optimization process is not typically completed with a single run due to various reasons, such as limited computing resources, incorrectly configured search algorithms, or large search spaces.
    Therefore, it is desirable for a system to track users' successive optimization processes and support their comprehensive analytic reasoning in the processes, eventually leading them to optimal results.
    We aim to design a flexible environment for representing the various hyperparameter sets in a scalable manner and tracking multiple optimization processes. Besides, we also aim to design a convenient environment for running a new process by making a tight connection with the backend framework.
\end{enumerate}

\section {\toolname: Visual Analytics for User-Driven Hyperparameter Optimization}
Based on the design goals identified in Section~\ref{sec:design goals}, we developed \toolname,\footnoteTendril a visual analytics for user-driven hyperparameter optimization of deep neural networks.
The interface of \toolname consists of a \controlPanel, an \optimizationOverview module, and an \mainModule module, as described below in detail.

\subsection{\controlPanel}
In order to support our design goal (\goalIterativeOpt), \toolname is designed to track and analyze multiple HyperOpt processes in a scalable manner. The \controlPanel (Fig.~\ref{fig:automlVisTeaser}(A)) allows users to control the HyperOpt processes, adjust those configurations (e.g., a search space, a search algorithm, and a computational budget), and run a new process with the revised configuration based on the analysis of previous results.

\begin{figure}[!t]
\centering
\includegraphics[width=0.9\columnwidth]{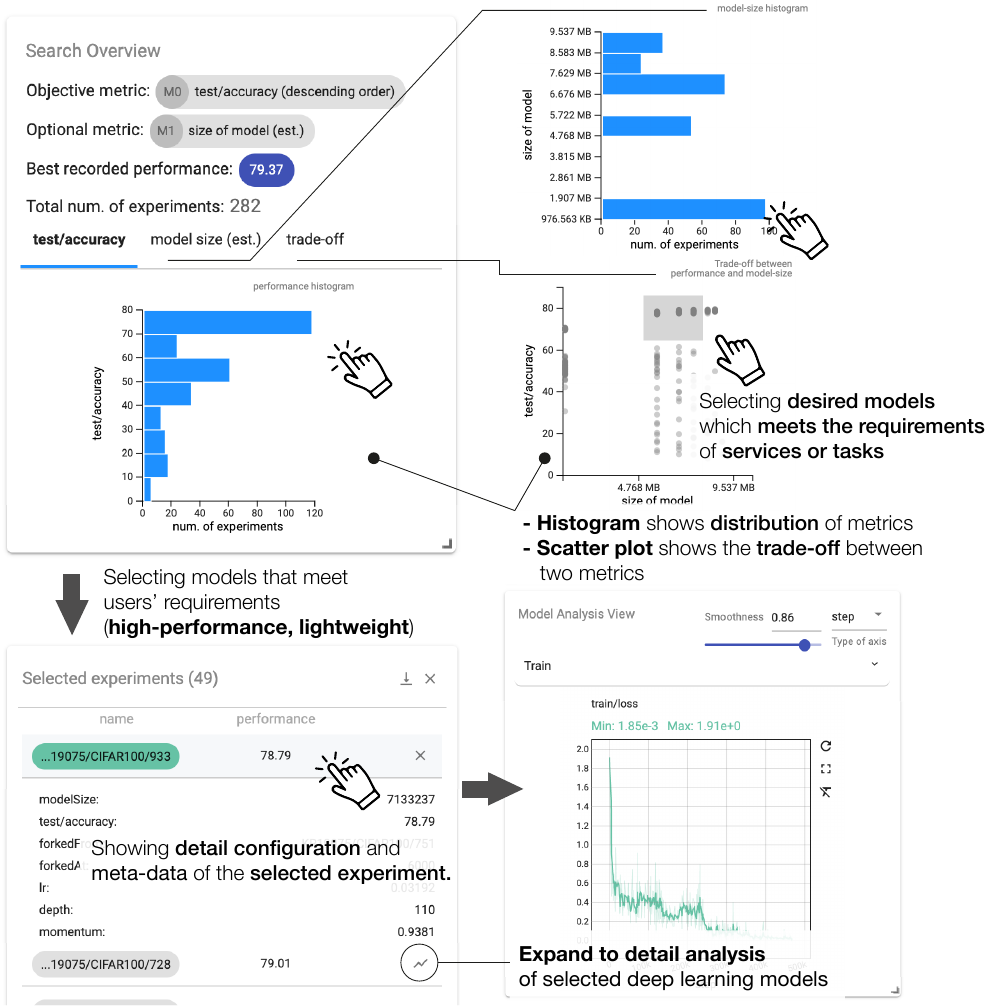}
\vspace*{-3mm}
\caption{The \optimizationOverview allows users to explore both the high-level statistics of optimization results and the distributions of deep learning models by two important criteria of the model size and prediction accuracy. When a user interacts with the histogram plots, it lists the target models and expand them to the detail analysis.}
\label{fig:optimization_overview}
\vspace{-18pt}
\end{figure}

\subsection{\optimizationOverview}
The \optimizationOverview (Fig.~\ref{fig:optimization_overview}) is designed to provide an overview of the optimization results of AutoML processes with high-level statistics, such as the number of experiments conducted in the processes, the highest model performance so far, as well as the distributions of the model performance along with additional metrics (e.g., the model size and the computing time for model inference).
In addition, \toolname supports a trade-off plot that helps users choose the best models satisfying their required metrics to support \goalSelectModel.
With these views, users can select the suitable models and include the models in the `\experimentTable' table to expand them to detailed analyses, such as checking detailed hyperparameter settings or examining the training process of the models.

\subsection{\mainModule}
The \mainModule summarizes the results of HyperOpt processes at three different levels of granularity, a \methodLevel, a \hyperparamLevel, and a \modelLevel, to support the essential tasks in optimizing hyperparameters of deep learning models.


\subsubsection{Search method-level analysis}
\begin{figure}[!t]
\centering
\includegraphics[width=0.85\columnwidth]{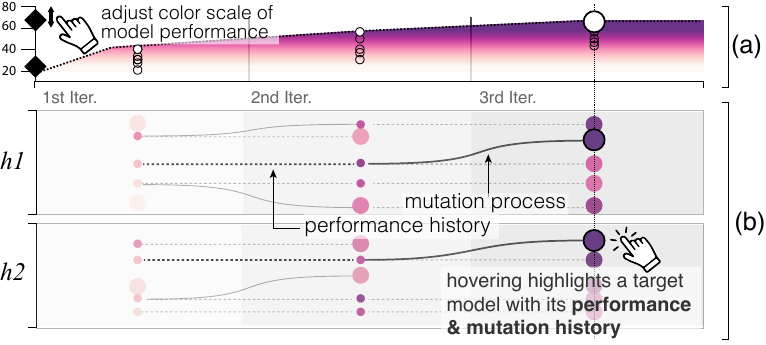}
\vspace*{-4mm}
\caption{
Visual encoding for \methodLevel analysis in \toolname: (a) monitoring model performance improvement and (b) exploring the history of each hyperparameter space.
}
\label{fig:exploration_plot_design}
\vspace{-18pt}
\end{figure}

When performing and analyzing a HyperOpt process, it is important to first determine whether the used black-box optimization algorithm worked properly on the target task (\needUnderstandMethod) before performing detailed analysis on effective hyperparameters and individual models.
While many visualization systems have so far devised effective interfaces to support HyperOpt tasks~\cite{golovin2017google, liaw2018tune, park2019visualhypertuner, tsirigotis2018orion, wang2019atmseer}, we have not found previous approaches which attempt visualizing the behaviors and patterns of various search algorithms that explore the given search spaces. To increase the transparency of AutoML, users should be able to understand and interpret the process of the black-box optimization.
To this end, \toolname provides interfaces from both macro and micro perspectives. 
The \explorationOverview is designed to support the diagnosis of HyperOpt algorithms and understanding of their behavior so that users can figure out whether the algorithm is properly configured (\goalUnderstandMethod).
The user interface is composed of two coordinated views (Fig.~\ref{fig:exploration_plot_design}) for (a) monitoring of performance improvement and (b) the exploration history of each hyperparameter search space, respectively.

First, the performance monitoring view (Fig.~\ref{fig:exploration_plot_design}(a)) visualizes the peak performance history of the created models in sequential order as the HyperOpt process progresses.
It allows users to check whether the process keeps making progress in improving the model performance.
In addition, \toolname uses a color encoding for the area plot with a scale of the model performance score.
By default, the color map corresponds to the range between the minimum and the maximum values, which users can interactively modify as well. 


The exploration history view (Fig.~\ref{fig:exploration_plot_design}(b)) is composed of multiple plots for each hyperparameter search space.
A single plot for a search space visualizes models with their hyperparameters.
Each model is visualized as a point, the x-axis presents the iteration index of the search algorithm, and the y-axis presents the value of its hyperparameter.
In addition, each represented model is encoded in darker colors with the higher performance so that users can confirm that the HyperOpt algorithms are exploring promising search space regions.
The color scale is updated by the user interactions on the performance monitoring view, as described before.

\begin{figure*}[!htbp]
\centering
\includegraphics[width=\textwidth]{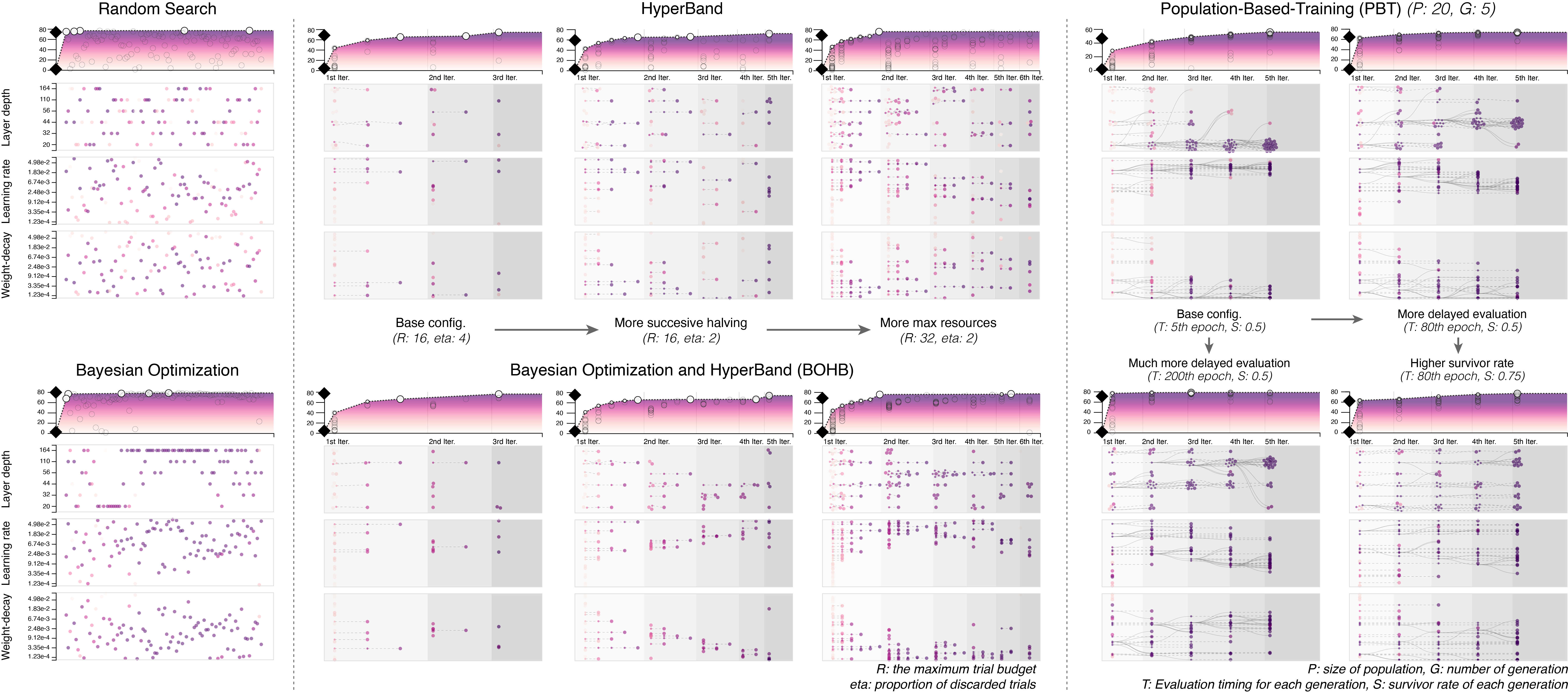}
\vspace*{-7mm}
\caption{
The exploration results of the five different HyperOpt methods with different configurations are visualized in the \explorationOverview: Random search (top-left), Bayesian optimization (bottom-left), HyperBand (top-middle), BOHB (bottom-middle), and PBT (right).
It shows that each method explores the same hyperparameter search spaces in a different manner, and the results vary with the configuration values of each method.
}
\label{fig:exploration_by_methods}
\vspace{-18pt}
\end{figure*}

Furthermore, the exploration history view is designed to assist users in understanding various types of search algorithms, by visualizing the exploration process.
We first categorized well-known search algorithms into four types and summarize each characteristic.
\begin{itemize}[leftmargin=*]
    \setlength\itemsep{-0.08em}
    \item Random search: Sample hyperparameters randomly.
    \item Sequential model-based~\cite{bergstra2011algorithms, shahriari2015taking}: Sample hyperparameters based on prior results.
    \item Bandit-based~\cite{falkner2018bohb, li2017hyperband}: Sample hyperparameters of configured sample size $R$ and evaluate after certain iterations (depending on the $R$ and $eta$). From the current set of models, successively discard the underperforming half at every evaluation step. Perform several successive halvings with the budget given by $R$.
    \item Population-based~\cite{jaderberg2017population}: Sample hyperparameters of configured population size $P$ and evaluate after certain iterations $T$. Discard the underperformers and keep the $k$ best performers given by survivor rate $S$ and population size $P$. Maintain the population size by copying the parameters of survived models and perturbing the hyperparameters. Repeat the process for $G$ generation steps.
\end{itemize}
By reviewing the algorithms, one may notice that bandit- and population-based methods also have a need to evaluate and compare between models along with sampling hyperparameters.
The difference between the two types of algorithms is that the bandit-based algorithm just discards the lower performers, but the population-based algorithm discards them and creates other models based on the surviving models.
In order to visualize these characteristics, \toolname first visualizes the history of model performance to represent the evaluation step in each algorithm's iterations with small points.
The survivors will have several points along the x-axis, and the small points for the performance history are connected with a single dashed line. The last point represents the final performance of the model, and it is visualized with a bigger point to distinguish from the preceding history.
Next, \toolname visualizes the mutation process of the population-based algorithm to represent how the algorithm creates new models from promising candidates of each generation.
Since the mutation process modifies the value of hyperparameters from a parent model, it is connected with a curved line and represented as a solid line to show that the point is a newly created model.
Lastly, since the algorithms initially sample hyperparameters in parallel by the characteristics, the generated models are aligned vertically. For the categorical type of hyperparameters (e.g., types of activation function), they can overlap in the visualization. We address this issue by applying repulsive force to the target coordinate so the points do not overlap.
For the evaluation and creation process of a single model, its history of the process is highlighted when the mouse cursor is placed on a point (bottom of Fig.~\ref{fig:exploration_plot_design}(b)).
In addition, by displaying the iteration index required by the specific algorithms as text and gradation (top and interior of the plot box in Fig.~\ref{fig:exploration_plot_design}, respectively), it is possible to intuitively and quickly understand how the algorithm searches the given search space and diagnose the algorithm's behavior.

We tested the \explorationOverview for five search algorithms with different configuration settings, as shown in Fig.~\ref{fig:exploration_by_methods}.
Random search and sequential model-based algorithms (left), which have no complex exploration process, reveal little search patterns in the given search spaces.
Bandit-based algorithms (middle) show that the promising models survived and the worse models were discarded.
Interestingly, although the `Bayesian optimization' does not reveal an apparent search pattern, the `BOHB', which is a combined algorithm of `Bayesian optimization' and `HyperBand', reveals a distinct search pattern.
On the other hand, the `PBT' reveals the distinct search patterns depending on the configuration values.

\subsubsection{Hyperparameter-level analysis}
One of the most important analytic needs found by the preliminary study (Section~\ref{sec:analytics_needs}) was to identify the hyperparameters with a significant impact on their model performances (\needEffectiveHyper).
In order to support such needs, \toolname provides both qualitative and quantitative analysis interfaces to identify effective hyperparameters (\goalEffectiveHyper).

\begin{figure*}[!htbp]
\centering
\includegraphics[width=\textwidth]{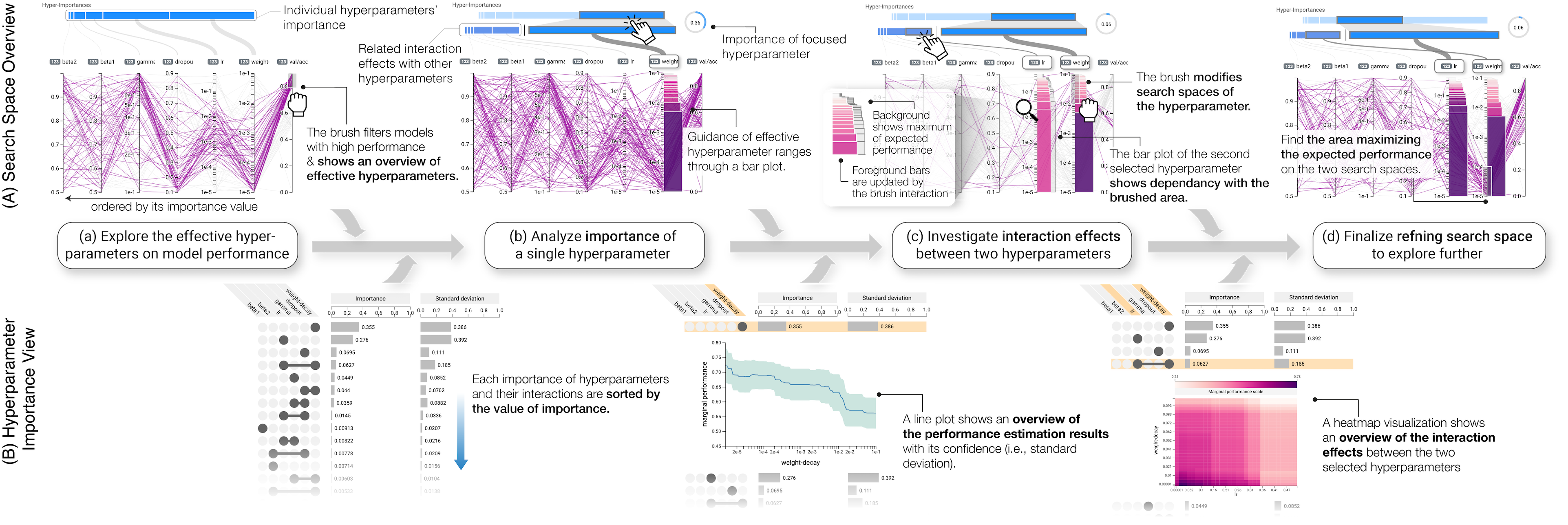}
\vspace*{-8mm}
\caption{
Illustration of visual exploration of (A) \searchSpaceView and (B) \hyperImpView to (a and b) identify effective hyperparameters, (c) analyze interaction effects between two selected hyperparameters, and (d) refine the search space to explore further.
}
\label{fig:hyperImportance_analysis}
\vspace{-18pt}
\end{figure*}

\textbf{\searchSpaceView}:
The \searchSpaceView (Fig.~\ref{fig:hyperImportance_analysis}(A)) utilizes a parallel coordinates plot~\cite{inselberg1987parallel} to present the overall search spaces and exploration results for the qualitative analysis support.
Based on this view, different combinations of hyperparameters are effectively visualized as high-dimensional vectors, together with a particular objective metric (e.g., a test accuracy) chosen by users. Also, its effective interaction capability can achieve our design goals to aid users in analyzing the effective hyperparameters (G1), filtering the desired models (G3), and refining the search space (G1) via brushing interactions. Hyperparameters are arranged in parallel along their corresponding axes, and the objective metric is placed in the last axis.
In the case of loading multiple HyperOpt processes (G4), which have different hyperparameter search spaces, the system flexibly adjusts the range of its corresponding axis using the minimum and the maximum values of each hyperparameter and objective metric.

In addition, to support the quantitative analysis, \toolname utilizes the functional-ANOVA (fANOVA) method~\cite{fanova2014efficient}, which measures the importance of hyperparameters based on the tested machine learning models.
In the machine learning research community, there have been numerous studies for assessing and quantifying the importance of hyperparameters~\cite{fawcett2016analysing, fanova2014efficient, jia2016qim, probst2018tunability, van2017empirical, van2018hyperparameter}.
Most studies quantify the importance of hyperparameters by building a performance estimation model that predicts the dependent variable (i.e., model performance) for the independent variables (i.e., hyperparameter configurations).
Among various approaches, we chose the fANOVA method since the method ensures linear-time performance in computing the importance, and it computes the importance of both single hyperparameter and interaction (i.e., joint) effects between them.

The top of parallel coordinates in the \searchSpaceView (Fig.~\ref{fig:hyperImportance_analysis}(A)) summarizes the results of the importance analysis. The visualization is composed of two layers of a bar plot for the importance of individual hyperparameters and the selected hyperparameter, respectively.
The width of each bar represents the relative importance between hyperparameters (Fig.~\ref{fig:hyperImportance_analysis}(a)), and each bar is ordered by its impacts from right to left. Each vertical curved line connects the corresponding hyperparameter axis of the parallel coordinates.
When users interact with each individual importance (b), the target hyperparameter is visualized in the second layer along with related interaction effects, and the estimated performance values in the search space region are visualized in the parallel coordinates as a bar chart.
Through the explicit guidance for the effective hyperparameters and their promising regions, \toolname can assist users in identifying the promising hyperparameters and in refining the search spaces for a subsequent HyperOpt process, by narrowing or widening them. Also, hyperparameters with low-performance expectations can be removed from the search spaces, thus effectively steering the search process.

The \searchSpaceView also provides an interface for investigating the interactions between other hyperparameters so that users can consider their relationships in modifying the search space (\goalEffectiveHyper).
If users interact with other hyperparameters in the second layer while a single hyperparameter is activated (Fig.~\ref{fig:hyperImportance_analysis}(c)), the \searchSpaceView summarizes the interaction effect between the two selected hyperparameters on the parallel coordinates, and users can investigate the effect of the brushed range on the other hyperparameter.
Fig.~\ref{fig:hyperImportance_analysis}(c) shows how a user can find a promising sub-space of the weight decay hyperparameter, considering the relationship with the learning rate hyperparameter.
When the user brushes with the range between $10^{-1}$ and $10^{-2}$ of the `weight decay' axis, the bar plot on the `learning rate' axis is updated to a low height, showing that the selected range could not produce a higher performance score.
When the user brushes with the range between $10^{-4}$ and $10^{-5}$, the bar plot is updated to a full height, showing that the range is good for the model performance (Fig.~\ref{fig:hyperImportance_analysis}(d)).


\textbf{\hyperImpView}:
While \searchSpaceView supports identifying the effective hyperparameters, visualizing details of the importance estimation on the view can overwhelm users. 
To provide the details, we designed the \hyperImpView, as shown in Fig.~\ref{fig:hyperImportance_analysis}(B).
The \hyperImpView utilizes a matrix-based visualization, which is designed to visualize set intersections~\cite{lex2014upset}, to represent the importance of both individual hyperparameters and interactions between the hyperparameters efficiently.
Each hyperparameter is placed in a column of the matrix, and each measured importance is listed by the value of importance so that users can quickly recognize and focus on the most important hyperparameters.
The corresponding hyperparameter for each row is visualized with a filled circle and is connected via a line if the row represents an interaction effect between hyperparameters.
The values of each importance and their confidence values are visualized next to the circles, and if a user interacts with a particular row, the details of the performance estimation result are visualized.
Fig.~\ref{fig:hyperImportance_analysis}(B) shows examples of the details for the importance estimation.
The bottom-middle of the figure presents the detail of the weight decay hyperparameter. Line and interval areas of the estimated model performance (y-axis) for the hyperparameter value (x-axis) are visualized. The green-colored interval shows the standard deviation of the estimation.
The bottom-right presents the detail of the interaction effect between the weight decay and learning rate hyperparameters. A heatmap visualization shows the estimated model performance according to the hyperparameter search spaces to provide an overall dependency between them.

\subsubsection{Model-level analysis}
\toolname allows users to switch from the result overview to in-depth analysis on demand so that they can evaluate the training process of the deep learning models (e.g., reviewing a learning curve for the diagnosis of under-fitting or over-fitting of models), supporting \goalSelectModel\xspace of our design goals. 
By filtering or selecting the models from each inter-linked components (B, C, and E1) in Fig~\ref{fig:automlVisTeaser}, the \modelAnalysisView visualizes the line plot of each model over iterations with several metrics users selected, as shown in Fig.~\ref{fig:automlVisTeaser}(D). It supports various interactions for the analysis of models such as aggregation of the moving average of a metric, area zooming, and others.

\section{Deployment to Industrial-scale Environment}
We have deployed \toolname on a cloud-based machine learning platform by NAVER called NSML~\cite{kim2018nsml, sung2017nsml}.
The NSML supports a HyperOpt framework called CHOPT~\cite{kim2018chopt}, and it provides various optimization methods, from simple random search to advanced algorithms~\cite{bergstra2011algorithms, falkner2018bohb, jaderberg2017population, li2017hyperband, shahriari2015taking}.
In addition, it provides the client with various APIs that can configure and run AutoML processes, and retrieve the recorded data.
Users who want to use the HyperOpt framework for their models can easily do so by adding only a few lines of code and a configuration file, which contains a metric to be optimized, hyperparameter spaces to be searched, methods for exploring the search spaces, and others.
Once an AutoML process is submitted to the AutoML agent, the users can utilize \toolname to explore the summarized results.
We note that the \toolname runs on top of the internal system, but it is designed to work with various AutoML frameworks.

For implementation details, each component was implemented with React.js\footnoteReact, and we used D3.js\footnoteDataDrivenDocument V5 to build visualization components.
The fANOVA method for estimating hyperparameter importance was implemented based on the official source code\footnoteFanova written in Python.
The generated model data by AutoML processes and the scalar data of the models from backend are passed to the interface with JSON format.
Meanwhile, the volume of scalar data generated from an industry-scale model is large in general, thus rendering performance is significantly reduced when users compare and analyze multiple models using the \modelAnalysisView. To mitigate this issue, we utilize a reservoir sampling~\cite{vitter1985random} technique to optimize the performance by fixing the size of the samples from the log data of each selected model.

\section{Case studies}
To demonstrate how \toolname can help users optimize their models and achieve our design goals, we conducted case studies in collaboration with two ML experts who have expertise in computer vision and natural language processing, respectively (denoted as P1 and P2).

\subsection{Understand, diagnose and refine AutoML method}
In this case, we illustrate how \toolname helps users understand and diagnose the executed HyperOpt algorithm and its configuration, which supports \goalUnderstandMethod, \goalSelectModel, and \goalIterativeOpt\xspace of our design goals.

\begin{figure}[!t]
\centering
\includegraphics[width=\columnwidth]{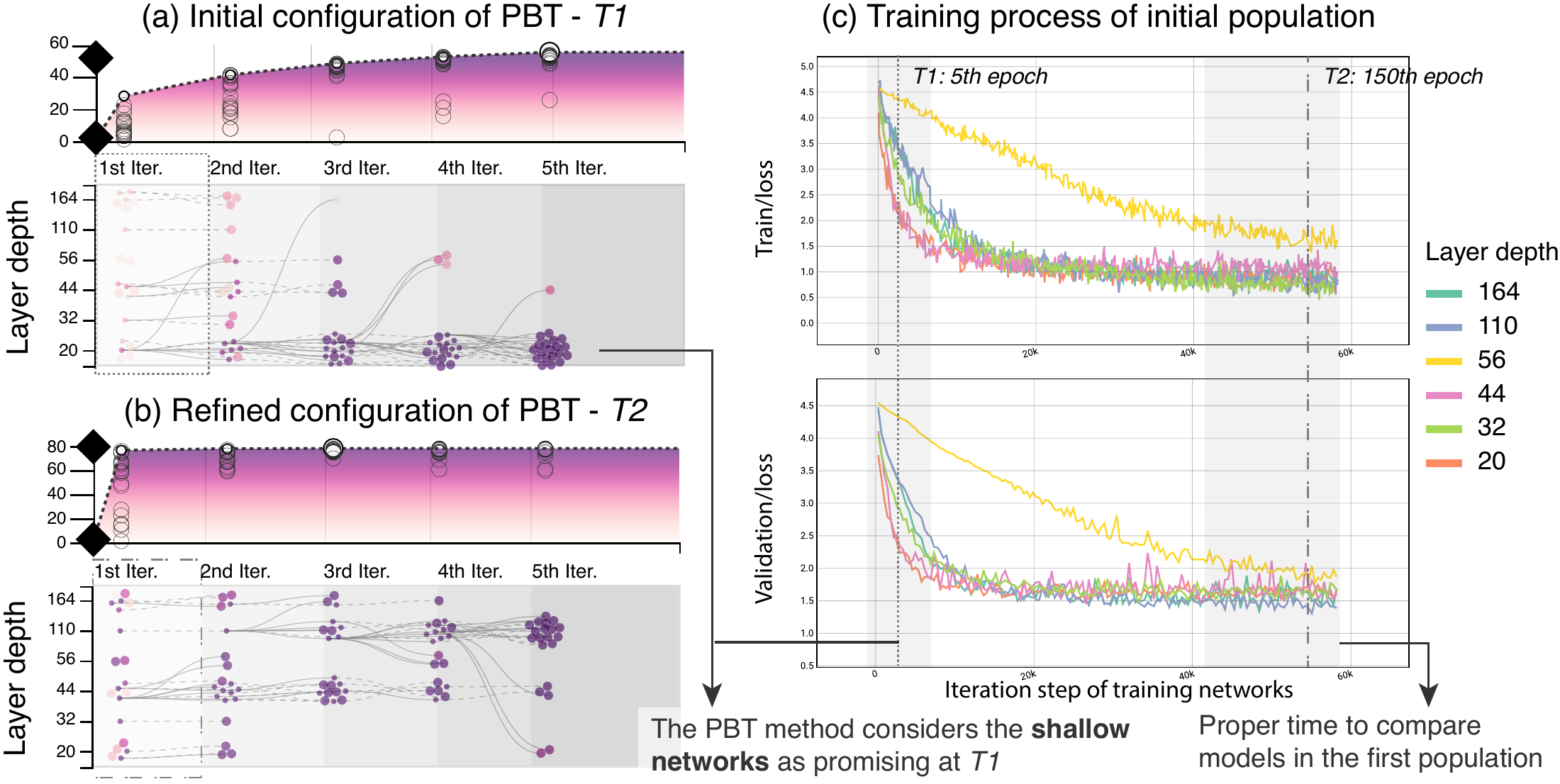}
\vspace*{-6mm}
\caption{
The \explorationOverview shows the results of two HyperOpt processes using the PBT method with different evaluation time $T$, configured by the user.
The \modelAnalysisView shows the training status of six models with different layer depths, which were sampled by the method at the initial stage. It shows that the method evaluates models' performance for each population at different times by the value of $T$.
}
\label{fig:loss comparison of diff layer}
\vspace{-18pt}
\end{figure}

P1, who is a machine learning engineer developing the image classification models, optimized the hyperparameters of the ResNet~\cite{he2016deep} model for CIFAR100 datasets. 
He first set the hyperparameter search spaces to optimize: layer depth, weight decay, and learning rate. 
With little prior knowledge of the methods of AutoML, he created a single AutoML process using the Population-Based Training (PBT)~\cite{jaderberg2017population} method as he heard that the method could effectively optimize deep learning models in a short amount of time.
Because he had no understanding of the method, he used the default configuration provided by the example code.
After the AutoML process was completed, he checked the results through \toolname and found that the recorded best performance score was 55.23\%, which is much lower than its known performance.
Through the \explorationOverview, he confirmed that the model's performance continually improved after a few iterations of the PBT algorithm, but he could not resolve why the performances were generally low (Fig.~\ref{fig:loss comparison of diff layer}(a)).

To find out the cause of the phenomenon, he analyzed the detailed behavior of the PBT method through the \explorationOverview. He observed that the method only focused on training networks with 20 layers and discarded the rest of the networks, which have deeper layers, as shown in Fig.~\ref{fig:loss comparison of diff layer}(a).
Following this observation, he analyzed the training status of networks with different layers, which were randomly sampled at the initial stage of the PBT method, through the \modelAnalysisView by interacting with them on the \explorationOverview (supporting \goalSelectModel).
The \modelAnalysisView (Fig.~\ref{fig:loss comparison of diff layer}(c)) revealed the fact that the training of each network was completed without saturating the loss function values and compared their performances by the AutoML method (T1 in Fig.~\ref{fig:loss comparison of diff layer}(c)).
Based on this finding, he was able to learn that performance evaluation time should be carefully configured to prevent the method from discarding models in a too early stage of training.
Utilizing the obtained insight, he refined the configuration of the PBT method by delaying the evaluation timing (as 150th epoch) and created a new AutoML process (supporting \goalUnderstandMethod\xspace and \goalIterativeOpt). He could obtain an accurate model of 78.07\% test accuracy score, about 23\% higher than of the initial process.

\subsection{Identify important hyperparameters and refine search spaces}
Next, we illustrate how \toolname helps users identify effective hyperparameters and refine those search spaces, supporting \goalEffectiveHyper\xspace and \goalIterativeOpt\xspace.


P2, who is working on sentiment analysis of customer reviews, optimized the hyperparameters of an LSTM~\cite{hochreiter1997long}-based model that classifies the positive and negative movie reviews in a major portal company.
She first set the hyperparameter search spaces to optimize: learning rate, weight decay, dropout, learning rate scheduler, two coefficients (i.e., betas, used for computing running averages of gradient) of Adam~\cite{kingma2014adam} optimizer.
She started an initial AutoML process with the Bayesian optimization method, which is set to explore 100 hyperparameter configurations.
After the initial process was finished, she first checked the best-recorded performance (89.9\%) and overall performance distribution in the \optimizationOverview.

Next, she wanted to identify the hyperparameters that have significant impacts on the validation accuracy, following the visual exploration illustrated in Fig.~\ref{fig:hyperImportance_analysis}.
She first brushed the top of the last axis of parallel coordinates in the \searchSpaceView to examine which regions of each hyperparameter are effective in the validation accuracy (Fig.~\ref{fig:hyperImportance_analysis}(a)).
She noticed that the lower values of the weight decay hyperparameter have impacts on the performance score, by observing a number of polylines at the bottom of the `weight decay' axis.
In addition, she could find out that the weight decay had the greatest influence on the model performance, by observing the list of the estimated importance boxes and the nearest axis of the last axis of the parallel coordinates.
To identify which regions are effective in the search space, she interacted with the weight decay hyperparameter box (Fig.~\ref{fig:hyperImportance_analysis}(b)).
After that, she could recognize that the effective values of the hyperparameter are under about $2\times10^{-3}$, by looking at the bar plot on the `weight decay' axis and the detailed estimation results in the \hyperImpView.
Following this, she interacted with the `learning rate' box located at the left of the `weight decay' box (Fig.~\ref{fig:hyperImportance_analysis}(c)) to examine the interaction effect with the highly correlated hyperparameter. Then she checked the overall dependency through the heatmap visualization in the \hyperImpView, and tried to find the area that maximizes the expected performance by brushing on the `weight decay' axis (Fig.~\ref{fig:hyperImportance_analysis}(c and d)).
After finding the effective area of the weight decay (between $10^{-3}$ and $10^{-4}$), she decided to perform a subsequent optimization process with more computational resources on the area.
In a similar manner, she performed analyses of effective regions on other hyperparameters and refined their search spaces.

After analyzing all the hyperparameters, she interacted with the \controlPanel and performed a new AutoML process with refined search spaces.
When the second process was complete, she noticed that the best performance score was 90.38\%, 0.48\% better than the previous process (supporting \goalEffectiveHyper\xspace and \goalIterativeOpt).
She also learned that the performance distribution is more stable than the initial process ($\mu$ = 0.871 (from 0.728), $\sigma$ = 0.059 (from 0.175)), implying that the refined search spaces yield generally good performance on her task. Satisfied with the results, she decided to stop the optimization process.

\section{User Study}
In this section, we evaluate the utility of \toolname.
In Section~\ref{quantitative study}, we summarize the usage behavior of \toolname with the interaction log data collected after the deployment.
In Section~\ref{qualitative study}, we describe various use cases of \toolname by the representative users and discuss the utility of \toolname through in-depth interviews.

\subsection{Interaction Log Analysis}\label{quantitative study}
\begin{figure}[!t]
\centering
\includegraphics[width=\columnwidth]{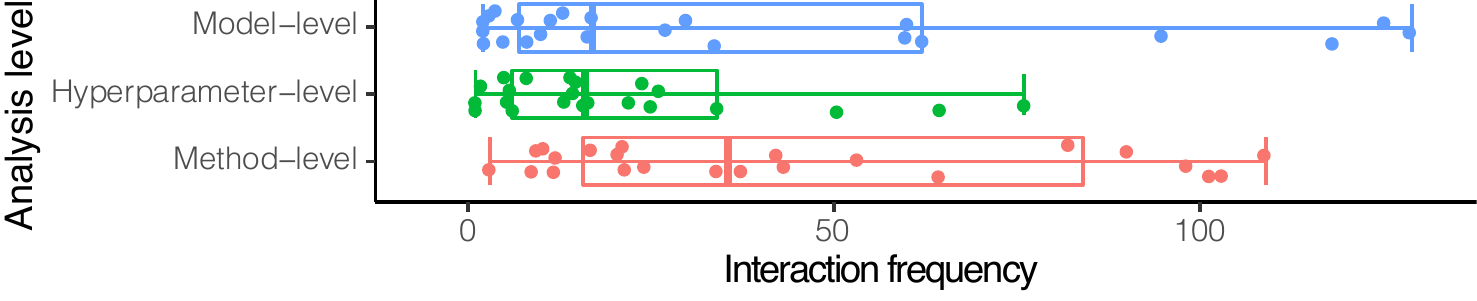}
\vspace*{-7mm}
\caption{
Interface usage distribution of \toolname with 32 users and 223 hyperparameter optimization processes.}
\label{fig:interface_usage}
\vspace{-18pt}
\end{figure}

In order to understand how users use the utilities of \toolname in general, we first collected and analyzed interaction logs for each interface.
We released an internal service of our \toolname from October 1, 2019, and we collected interaction logs from November 1, 2019, to March 1, 2020.
There was a total of 32 users during the period, and the interaction logs for 223 AutoML processes were collected.
We categorized the collected logs according to each analysis level of granularity and summarized the interface usages with the interaction frequency, as shown in Fig.~\ref{fig:interface_usage}.
Based on the analysis, we found out that the distribution of interaction frequency is evenly distributed by each analysis level.
This result means that users have various types of analytic needs in performing the HyperOpt task, as identified in Section~\ref{sec:analytics_needs}, and that \toolname has properly supported their needs.


\subsection{Use Cases with \toolname and Feedback}\label{qualitative study}
To better understand how users performed HyperOpt tasks for their problems with \toolname, we conducted in-depth interviews with engineers and scientists who actively used our tools based on the log analysis we performed. We summarize key findings and feedback from these studies to highlight \toolname's benefits.

\subsubsection{Participants and Study Protocol}
We selected and recruited the three most active users of \toolname with a different team, domain, and their usage behavior:

\par
\begingroup
\leftskip0.5em
\rightskip\leftskip
\textit{User A} is a machine learning engineer who has expertise in natural language processing.
He works with the news team to develop classification models for malicious comments in news articles.
\par
\endgroup

\par
\begingroup
\leftskip0.5em
\rightskip\leftskip
\textit{User B} is a research engineer who has expertise in computer vision. He is developing a model that retrieves similar images or blog postings on the web, based on given images.
He is interested in not only effectively tuning the model performance, but also optimizing the model size that could be mounted in edge devices.
\par
\endgroup

\par
\begingroup
\leftskip0.5em
\rightskip\leftskip
\textit{User C} is a research scientist who has expertise in machine learning and AutoML. 
Unlike A and B, he studies and develops AutoML frameworks, including hyperparameter optimization, neural architecture search, and others. He is interested in the differences in the behavior of various optimization algorithms.
\par
\endgroup

We had a 60-minute session with each of the three participants.
For the first 20 minutes, we asked them a few questions about their tasks, typical workflows of their HyperOpt processes, and the main intent of analysis when using \toolname. 
We then asked them to revisit the process they performed with \toolname and describe the sequence of interactions when analyzing the results while thinking aloud.

\subsubsection{Main Findings and Feedback}
We summarized the main findings and feedback from the interviews into the following criteria, by highlighting how the \toolname helps to perform their optimization processes and achieves our design goals.


\textbf{Diagnose AutoML and algorithm configuration.}
Users A and C insensitively used the PBT algorithm to perform HyperOpt processes. 
Both commented that \toolname helped them understand how the algorithm works on their tasks and diagnose their configurations, satisfying \needUnderstandMethod\xspace and \goalUnderstandMethod.
User A said that ``Although we have never used the PBT method before, \explorationOverview helped to build a mental model of how the method explores the search spaces.
We initially set the algorithm configuration with a default value but observed that the process converges too quickly to a particular point without performance improvement due to the low survivor rate of each generation.
Then we expanded the diversity of the population by increasing the survivor rate and could find better models.
The visualization can help to analyze the trade-offs between exploration and exploitation and to determine which configuration is appropriate for our task." 

\begin{figure}[!t]
\centering
\includegraphics[width=0.9\columnwidth]{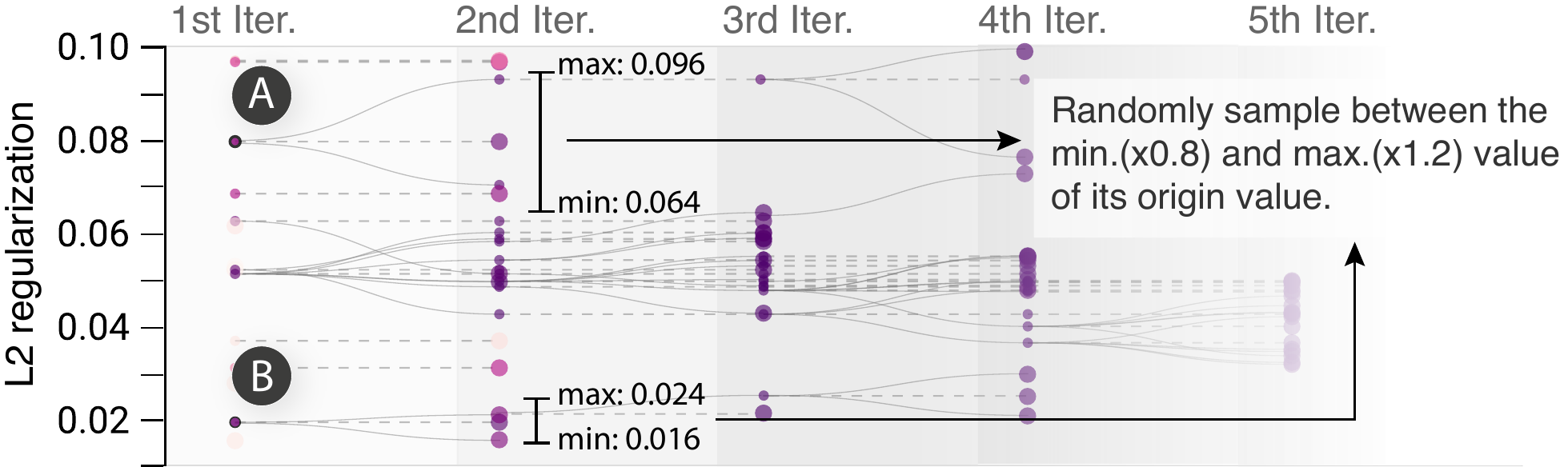}
\vspace*{-3mm}
\caption{
The use case of the \explorationOverview to diagnose an AutoML method. It shows that the PBT method can have a bias for low-value regions when exploring search space if there is no proper scaling for perturbation in its implementation.
}
\label{fig:pbt_perturbation_bias}
\vspace{-18pt}
\end{figure}

User C commented that the \toolname could assist AutoML algorithm developers in identifying bugs in the implementation of AutoML methods.
Fig.~\ref{fig:pbt_perturbation_bias} shows the results of the PBT optimization method.
The \explorationOverview reveals that the method tends to explore the low-value regions in the search space, and the perturbation scale in the high-value region (Fig.~\ref{fig:pbt_perturbation_bias}(A)) is greater than that of the lower values (B).
Through the visualization, User C was able to quickly recognize the strange behaviors of the perturbation and identify the bug in the current implementation of the AutoML algorithm, by incorporating his domain knowledge into the analysis~\cite{tam2016analysis}.
In the AutoML framework, the PBT method copies promising model's parameters and randomly perturbs its hyperparameter with noise (usually by a factor between 1.2 and 0.8 of the original value) in the population.
However, the current implementation performed perturbation from the promising hyperparameter value without considering the scaling factor.
In this case, hyperparameters of smaller values tend to have smaller perturbation range compared to hyperparameters of larger values, resulting in the method to be biased toward the low-value regions when exploring the search space.
User C said that ``It was difficult to identify problems in algorithm's implementation that did not produce an error using the only console of standard-based output, but the \explorationOverview helped to identify implemented algorithm's abnormal behavior by displaying the search history and its tendency."


\textbf{Identify effective hyperparameters.}
All the participants appreciated the interactive and iterative capabilities for the HyperOpt process based on the guidance of effective hyperparameters (\needEffectiveHyper\xspace, \goalEffectiveHyper\xspace, and \goalIterativeOpt).
They believed such interaction could improve an AutoML process. 
They commented that ``Human involvement with prior knowledge and observation sometimes can be more efficient than optimization algorithms, especially when there are a large search space and limited computational budget."
User A said that ``Even if there are enough time and computational budget, it takes a long time to try all the cases and check the results.
So the optimization process can be more effective if we can focus on a few effective hyperparameters. The \hyperparamLevel analysis view helps to refine search spaces by suggesting hyperparameters that should be dropped or explored more in a subsequent optimization process, and it helps engage the optimization task by providing the importance quantitatively."
In addition, they valued the complementary interactions with the \hyperImpView and \searchSpaceView.
User B said that ``Effective and ineffective hyperparameters could be observed in the \searchSpaceView, but it can be verified more in the \hyperImpView, enabling more accurate reasoning for search space refinement."

\textbf{Select and validate target models.}
All users appreciated the fact that visual exploration provided the ability to quickly select interesting models out of a large number of models (\needSelectModel\xspace and \goalSelectModel).
User B, who is closely working in real services, actively used and appreciated the \optimizationOverview and \modelAnalysisView to validate his models for the service deployment. 
He said, ``The trade-off plot in the \optimizationOverview helps to filter models that do not satisfy the required network size for our service. 
Also, the exploration flow from the overview to the in-depth analysis of filtered models is intuitive, and it was an essential part of the final validation since we usually reported a dozen of validation-related metrics in each iteration during model training."
Meanwhile, Users A and C commented that the \modelAnalysisView helps to validate and determine the proper evaluation time in each exploration loop of HyperOpt algorithms.
User C said, ``Because sometimes the slope of the loss function value can vary depending on the hyperparameter of the deep learning model, setting the proper validation time is important in using AutoML. Therefore, it is necessary to check the value of loss function through the \modelAnalysisView and set the appropriate validation time so that the early stopping method can be operated without biases to specific hyperparameters."


\section{Discussions and Future Work}
\textbf{Scalability of visualization.}
Although there may be some scalability issues on parallel coordinates in \toolname, the average number of hyperparameters used by users in the real industrial environment was found to be around 5 with a maximum of 16. It was partly due to users' limited cognitive load as well as limited computational budget and time. 
Therefore, we believe that the scalability of parallel coordinates in terms of the number of dimensions is acceptable for most of the real-world HyperOpt tasks in a cloud-based machine learning platform.
Another scalability issue is caused by the increasing number of tested hyperparameter settings, each of which is drawn as a polyline in parallel coordinates. A technique for bundling polylines based on the hyperparameter importance will be studied in future work.
\newline
\textbf{Restricted hyperparameter importance estimation.}
Finding effective hyperparameters is one of the most important needs in HyperOpt tasks. However, in the PBT method, critical limitations exist for the fANOVA approach we employed because the final model can utilize multiple sets of hyperparameters during the HyperOpt process, resulting in the incorrect results in the variable analysis. A further study on the methodology would be helpful in handle the incorrect results.

\section{Conclusion}
We presented \toolname, a visual analytics system that supports the analysis and steering of the hyperparameter optimization process for deep neural networks.
We conducted preliminary interviews with ten machine learning practitioners across various domains to identify their analytic needs.
Based on the interviews, we distilled four main design goals: (1) guidance for effective hyperparameters; (2) understanding and diagnosing various AutoML methods; (3) reasoning and filtering target models; (4) supporting iterative optimization processes.
We then proposed a visual analytics system allowing users to examine multiple AutoML processes and analyze the results at three levels of granularity: search method-, hyperparameter-, and \modelLevel analysis.
\toolname has been deployed on top of a machine learning platform in a major software company.
We presented a user study with real-world users in the platform and use cases of how \toolname can be utilized with different applications. Our results showed that users appreciated the utility of \toolname in refining the search space and the method configuration of AutoML.


\footnotesize{\acknowledgments{This work was partly supported by Basic Science Research Program through the National Research Foundation of Korea (NRF) funded by the Ministry of Science, ICT \& Future Planning (2019R1A2C4070420) and by Korea Electric Power Corporation (Grant number:R18XA05).
}}


\newpage

\bibliographystyle{abbrv}
\bibliography{template}

\newpage

\end{document}